# Quasi-metallic conductivity in mammalian pigment eumelanin thin films after simple thermal vacuum annealing.

Ludovico Migliaccio,[1] Paola Manini,[1] Davide Altamura,[2] Cinzia Giannini,[2] Paolo Tassini,[2,a] Maria Grazia Maglione,[2] Carla Minarini,[2] and Alessandro Pezzella[1,4,5,b]

[1]Department of Chemical Sciences, University of Naples "Federico II", Via Cintia 4, IT-80126 Naples, Italy.
[2]Istituto di Cristallografia (IC), CNR, via Amendola 122/O, IT-70126 Bari, Italy.
[3]Laboratory of Nanomaterials and Devices (SSPT-PROMAS-NANO), ENEA C. R. Portici, Piazzale Enrico Fermi 1, Località Granatello, IT-80055 Portici (NA), Italy.
[4]Institute for Polymers, Composites and Biomaterials (IPCB), CNR, Via Campi Flegrei 34, IT-80078 Pozzuoli (Na), Italy.
[5]National Interuniversity Consortium of Materials Science and Technology (INSTM), Via G. Giusti, 9, 50121 Florence, Italy

Corresponding authors: [a]paolo.tassini@enea.it; [b]alessandro.pezzella@unina.it.

## Abstract

*Melanin* denotes a variety of mammalian pigments, including the dark electrically conductive eumelanin and the reddish, sulphur-containing, pheomelanin. Organic (bio)electronics is showing increasing interests in eumelanin exploitation, e.g. for bio-interfaces, but the low conductivity of the material is limiting the development of eumelanin-based devices. Here, for the first time we report an abrupt increase of the eumelanin electrical conductivity, presenting the highest value reported to date of 318 S/cm, obtained integrating together the knowledge of the eumelanin chemical properties with simple thermal annealing in vacuum of the material thin films, unveiling the actual electronic nature of its conduction.

## Introduction

In the 1974, McGinness *et al.* reported the first experimental evidence of the semiconducting behaviour of the eumelanin,[1] the polyindolic pigment responsible, inter alia, of the dark-brown pigmentation of the mammalian (including human) skin, hair and iris. The study followed a pioneering suggestion by Pullman and Pullman [2] on the possible existence of energy bands associated with a not-localized empty molecular orbital within an infinite eumelanin polymer chain acting as an one-dimensional semiconductor.
Since then, the charge transport properties of this challenging materials class were extensively studied, [3] and particularly in the recent years, after the renewed interest in the topic, because of the prospect of eumelanin applications in organic (bio)electronics.[4] To date, eumelanin conductivity is reported in the range [5,6] $10^{-13}$-$10^{-5}$ S/cm, largely depending on the measuring conditions, and especially on the presence of humidity in the measuring environment.[7] For valuable applications, higher conductivity values are needed yet, thus several studies explored the integration of the eumelanin with other more conductive materials, [8-10] but strongly affecting its chemistry, or exploiting severe modifications of eumelanin-like materials to gain a graphene-like material, as for example by pyrolytic treatment of polydopamine under hydrogen or argon atmosphere. [11, 12]

Although the mechanisms of charge transport in eumelanin are still not fully clear, several evidences are concurring to sustain its hybrid ionic-electronic behaviour, [13, 14] where the electronic contribution depends on the presence, extent and the redox properties[13] of the delocalized aromatic systems, while the ionic part is largely dictated by the hydration level of the material[14] (i.e. the humidity in the measuring environment).

Basing on the concurring evidences disclosing the correlation between the chemical-physical properties of the eumelanin and the polyindole π-system stacking, as well as the packing of molecular constituents within the material, [15, 16] we speculated about the modulation of the electronic conductivity[17, 18] by acting on the polyindole packing in eumelanin thin films. This is bringing us, here, for the first time in our knowledge, to report the preparation and characterization of eumelanin thin films showing the highest conductivity values of this material up to 318 S/cm. Conductive eumelanin films were prepared via the preliminary oxidative polymerization of the solid state form of the 5,6-dihydroxyindole (DHI, the ultimate monomer precursor in the formation pathways of natural and synthetic eumelanin (Figure S1)),[19] and then by thermal annealing of the material films, at temperatures no higher than 600°C and under high vacuum conditions. We name the obtained material as High Vacuum Annealed Eumelanin, HVAE.

**Samples preparation**

All the commercially available reagents and materials were used as received. All the solvents were analytical grade quality. The DHI was prepared according to a reported procedure.[19]

The samples were prepared on quartz substrates (dimensions 15 mm X 6 mm X 1.2 mm), cleaned by sonication in a solution of detergent Borer Chemie AG Deconex 12PA® in deionized water (18 MΩ·cm) at 70°C for 30 min, and rinsed in acetone and then in isopropanol for 15 min each sequentially. A concentrated solution of DHI in methanol-ethyl acetate (1:1 v/v) (50 mg/mL) was prepared, filtered through a 0.2 μm Whatman membrane before deposition; on each sample, 15 μL of this solution were applied. Thin films were obtained by spin coating, using a Laurell WS-650MZ23NPP/LITE coater, with the spinning recipe acceleration 2000 rpm/s, speed 3500 rpm, duration 30 s; the samples were then dried at 90°C for 30 min in oven in air, the thicknesses of the resulting films were 230 nm +/- 10 nm, measured using a stylus profilometer KLA Tencor P-10.

The eumelanin formation was obtained by the oxidation of the DHI films thanks to the Ammonia-Induced Solid State Polymerization (AISSP) method, a recently developed solid state protocol: [19,20] each sample was exposed for 12 hours to an oxidizing atmosphere made of oxygen and ammonia vapours at controlled temperature (25°C), produced by the equilibrium of the air with an ammonia solution (5% $NH_3$ in $H_2O$) in a sealed chamber at 1 bar pressure. The thicknesses of the resulting films were 260 nm +/- 6 nm. Films showed the typical dark brown colour of the eumelanin, presenting flat surfaces (Figure S2, Figure S3, Table S1; surfaces roughness images were taken using a Taylor Hobson® CCI-HD non contact 3D Optical Profilometer with thin & thick film measurement capability; films' roughness was estimated as a Root Mean Square (RMS) value from several scans on each sample). This material is here named DHI-eumelanin, to distinguish it from the starting DHI, and from the final HVAE.

The eumelanin films were then annealed at different controlled temperatures (230°C, 300°C, 450°C and 600°C, +/- 1°C for each value) in high vacuum conditions ($10^{-6}$ mbar); some samples were annealed at various time lengths (from 30 min up to 6 hours). The processes were performed in a dedicated high vacuum chamber using a turbomolecular pump to obtain the vacuum level, and doing preliminary leak detection and samples temperature verifications. The mean thickness of the HVAE films was dependent on the annealing conditions, with the smallest values down to 110 nm +/- 2 nm for the processes at 600°C longer than 1 hour (Figure S5).

**Characterizations and comments**

The chosen annealing temperatures are well below the values reported as the starting temperature for the degradation[21] and/or the carbonization processes in similar materials[22], but includes a significant part of the eumelanin mass loss region, as shown by thermogravimetric analysis (TGA) performed under not oxidizing atmosphere (Figure S4), using a Perkin–Elmer Pyris 1 thermogravimetric analyser. Moreover, applied temperatures include the complete loss of both weakly and strongly bound water [5, 21, 23], as well as the loss of $CO_2$ from carboxyl groups in DHI-melanin (thermal decarboxylation).[24] Indeed, TGA data under not oxidizing conditions indicate that mass loss is nearly completed at 800°C, suggesting that, beyond the loss of volatile elements, no modification of the molecular backbone occurs at 600°C. Instead, a complete different picture is obtained in presence of oxygen, which critically affects the stability of the material (Figure S4).

Morphology and surface analysis of the materials at the different stages of the process revealed a nearly unmodified roughness, passing from the starting DHI films to the HVAE films (Figure S3) (using the definition of the roughness according to the standard ISO 25178; DHI roughness = 6.45 nm; DHI Eumelanin roughness = 6.52 nm; HVAE roughness = 6.58 nm), while, the thicknesses suffered a significant decrease in function of the annealing temperature from 260 nm to 109 nm in the case of the sample treated at 600°C (Figure S5). This was expected because of the known tendency of the eumelanin to loss labile carboxylic groups[19, 21, 24] and on the possible loss of low molecular weight components embedded in the material layers (i.e. small species arising by oxidative ring fission of DHI during its melanization).[25]

Scanning electronic microscopy (SEM) inspection (using a SEM Zeiss Leo 1530 Gemini) confirmed the retaining of the high quality morphology in the HVAE films (Figure S6), showing an uniform surface of this material.

UV-Vis inspection was carried out using a Perkin-Elmer Lambda 900 spectrophotometer. Samples were observed at the different process steps (Figure 1) to record the spectra. It can be seen an evident increase in the absorption coefficients in nearly the entire UV-Vis range, passing from the DHI to the DHI-eumelanin and to the HVAE. This phenomenon is associated to the increase of both the delocalization of the aromatic systems and their π-stacking interactions,[15, 16] that suggest the actual increase of the extension and of the filling factor[16, 23] for the delocalized aromatic systems of the material backbone, in particular happening after the thermal annealing in vacuum: i.e. this reorganization results in an overlap of the π-electronic density of the adjacent packed chains and the delocalization of their electronic wave-functions.[26]

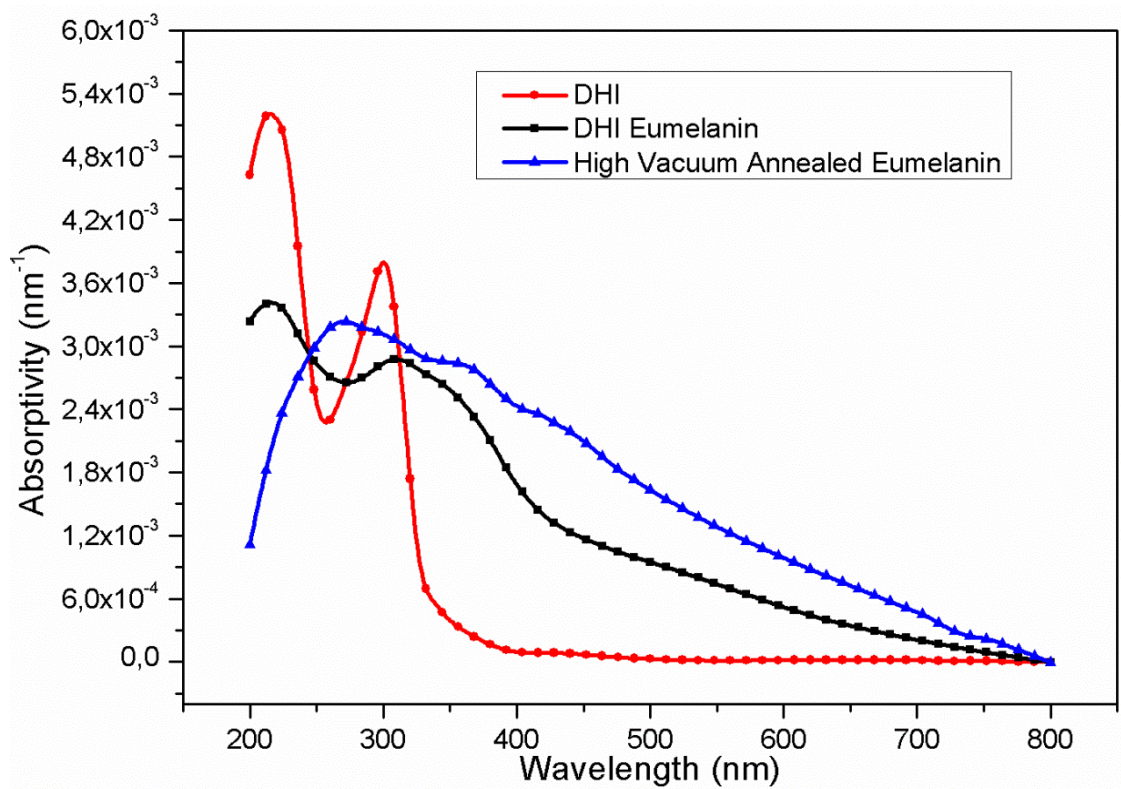

*Figure 1. UV-Vis absorptivity (percent absorbance / film thickness) of the films at the different process stages: (red, circles) DHI thin film; (black, squares) eumelanin film after AISSP; (blue, triangles) HVAE film after thermal annealing in vacuum (600°C; 2 h; $10^{-6}$ mbar).*

Strong supports to the picture of a structural reorganization and an enhanced packing order[27, 28,29] of the molecular constituents within the eumelanin films (made possible by the concomitant loss of labile and low molecular weight components[24] and the clustering of the longer polyindole chains (a pictorial representation of these mechanisms is shown in Figure 2)), were further given by the retaining, in the annealed films, of the typical eumelanin signature observed in the electron paramagnetic resonance (EPR) spectrum[5, 19] (measured using an X-band (9 GHz) Bruker Elexys E-500 spectrometer, equipped with a superhigh sensitivity probe head), in the FTIR analysis[30] (using a Thermo Fischer Scientific Nicolet 6700 FTIR to determine the attenuated total reflectance (ATR) spectra of the samples, with a resolution of 4 $cm^{-1}$ and 16 scans averaged for each spectrum in a range between 4000–650 $cm^{-1}$), in the Raman spectroscopy[21, 31] (Renishaw inVia 2 Raman microscope (532 nm), which uses a microscope to focus a laser source onto specific areas of a sample, then the light scattered off the surface of the sample is collected and directed to a Raman spectrometer), and in the MALDI-MS[20] analysis (positive reflectron MALDI and LDI spectra were recorded on a Sciex 4800 MALDI ToF/ToF instrument) (Figures S7 to S10).

Without entering into the details of the Raman spectra (Figure S9), it is worth to note here how the comparison of the profiles before and after the annealing reveals, in agreement with the loss of carboxylic groups and possible pyrrolic acids, a relative reduction of the G band (the range of 1600 $cm^{-1}$) following the reduction of O and N contribution.

Consistent information is provided by the FTIR spectra of eumelanin and HVAE films (Figure S8) too, highlighting in particular the drastic decrease of signals associated to C=O stretching (1620 cm$^{-1}$) and to the water (3200 cm$^{-1}$).[30]

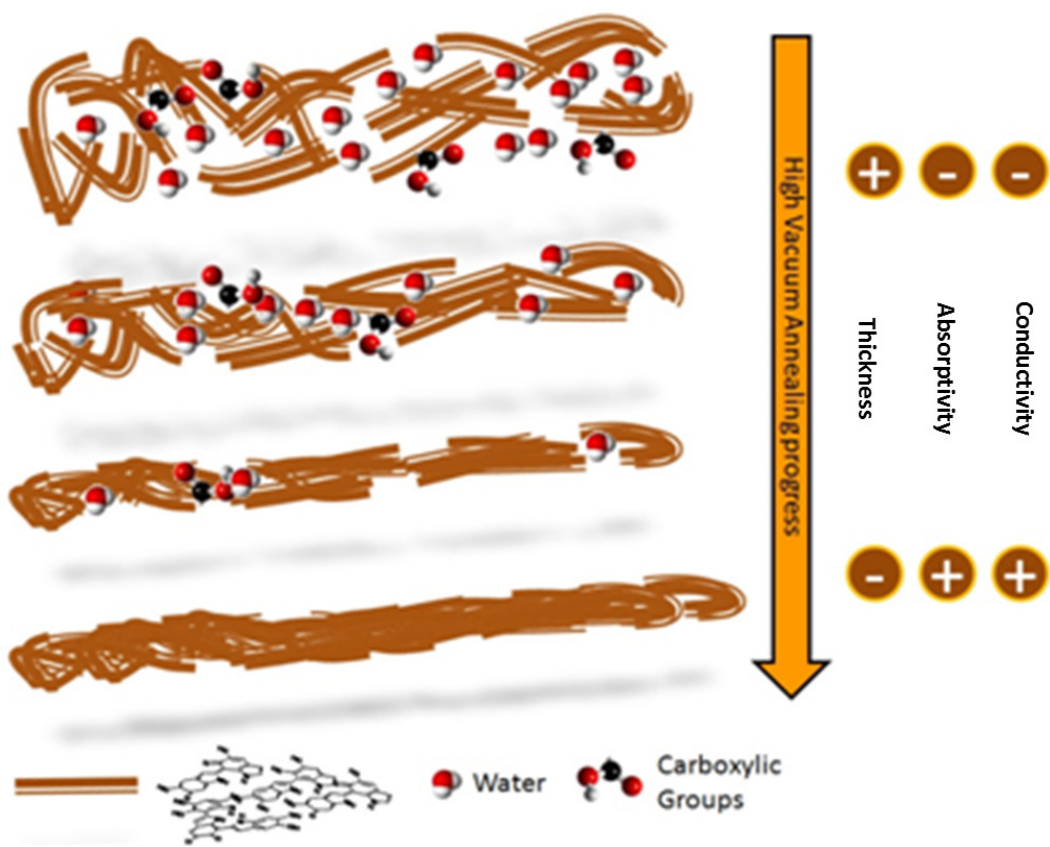

Figure 2. Pictorial model of the polyindole packing evolution during high vacuum annealing. Water molecules and carboxylic groups are evidenced, to show their reduction in the material as the process temperature increases.

Finally, a direct support to the packing evolution hypothesis comes from 2D GIWAXS patterns (Figure 3), where the different anisotropy degree of the intensity distribution along the diffraction rings indicates an increased orientation degree after the vacuum thermal treatment is operated. In particular, the HVAE film (Figure 3A) features a diffraction intensity definitely concentrated along the $Q_z$ axis, i.e. perpendicularly to the sample surface, denoting a preferred orientation of the diffracting planes parallel to the film surface. On the other hand, the Eumelanin film (Figure 3B) features a weak diffraction intensity evenly distributed along the azimuth of a broad diffraction ring, indicating low crystallinity and random orientation of the molecules. The 1D radial cuts extracted from the GIWAXS maps along the out-of-plane (Figure 3C) and in-plane (Figure 3D) directions show indeed a clear difference between the two directions in the case of the HVAE film: a peak asymmetry in the out-of-plane direction reveals indeed a diffraction contribution of the oriented molecules appearing as a shoulder at $q$ = 1.85 Å, at the side of the main peak at $q$ = 1.56 Å which is ascribed to the substrate and is in turn the only scattering contribution in the in-plane cut. The shoulder in the out-of-plane direction is a clear signature of the formation of a well oriented stack, compatible with the expected supramolecular structure with 3.4 Å periodicity.[32]

On the contrary, no difference between the diffraction intensities in the two directions is recognized in the case of the Eumelanin film (so that a 5.5 additional scale factor has been applied in Figure 3C for the sake of clarity in the comparison).

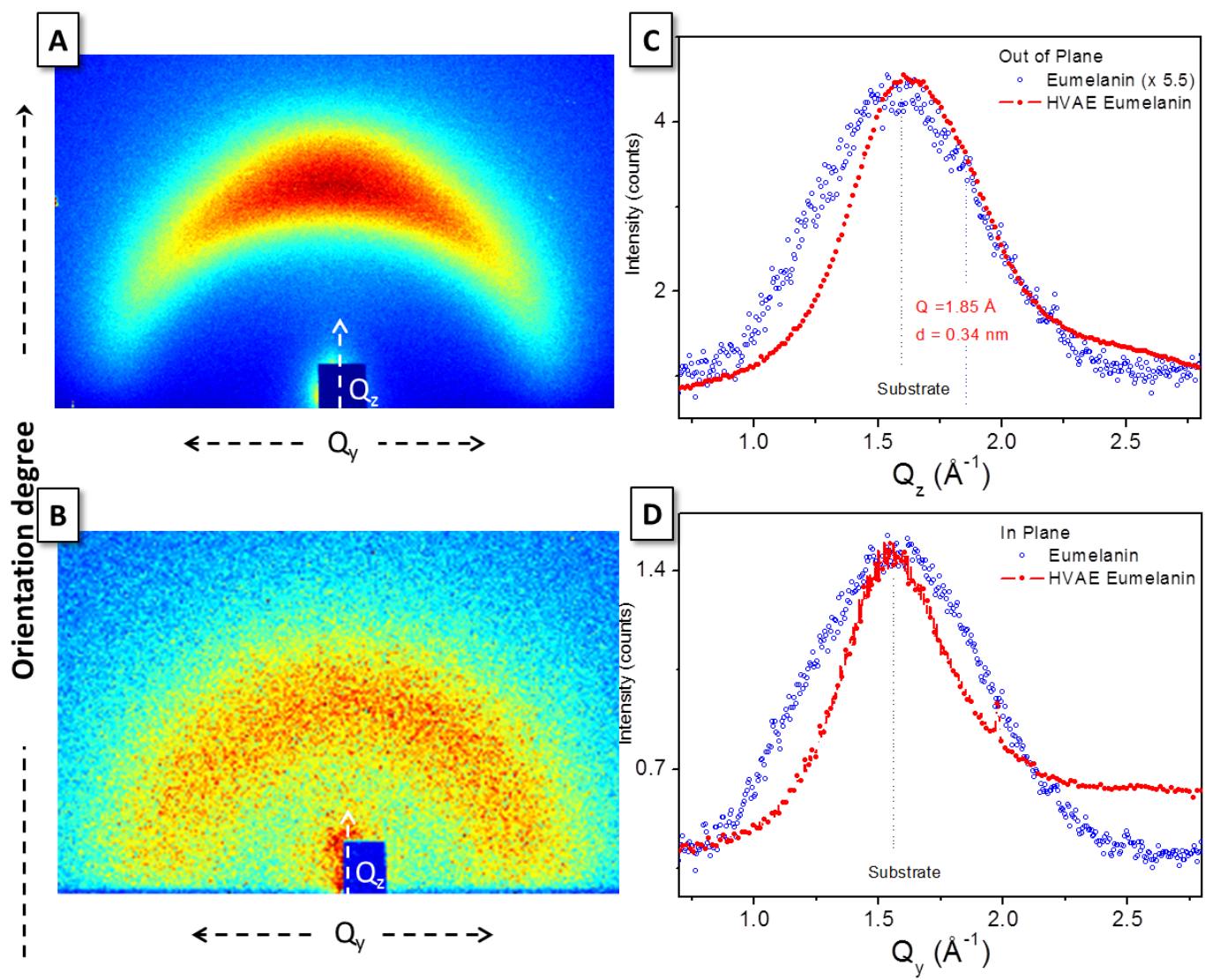

*Figure 3. GIWAXS 2D patterns of HVAE films processed at (A) 600°C (2h treatment) and (B) eumelanin thin film. 1D radial cuts along (C) the out-of-plane and (D) the in-plane directions, obtained from the 2D maps in (A) and (B).*

The electrical properties of the various materials were measured using a four probes system[33] Napson RESISTAGE RG-80 and a power supply source meter Keithley 2410 for the different ranges of the conductivity, working in air at room temperature. The samples conductivity vs. the annealing temperature and vs. the duration of the processes are shown in Figure 4. After the vacuum annealing, the conductivity of the eumelanin films featured a remarkable increase, up to over 9 orders of magnitude, passing from around $10^{-7}$ S/cm for the DHI and DHI-eumelanin films, up to an unprecedented value of 318 S/cm for the material processed at 600°C for 2 hours, and anyway obtaining values larger than $10^2$ S/cm for all the samples processed at 600°C (Figure 4 inset).

This record result is not a humidity response effect, as the data acquisitions were performed in few tens of seconds for each sample, with no variation of the ambient relative humidity, so suggesting that actual nature of the involved charge carriers is electronic.

Current-voltage measurements performed before and after the exposition of the films to water or acidic conditions conclusively ruled out any conductivity increase with the water content of the film. Immersion of the films in deionised water results in a marked decrease of the conductivity, also associated to a deterioration of the surface smoothness (Figure S11 and Table S2). Reduction of the conductivity is even more pronounced when films are exposed to acidic solutions[19] (Figure S12 and Table S3). Notably, the films appear moderately stable under accelerated ageing (Table S4), but stability is lost if the film were previously immersed in water (Table S2). In light of known literature,[14, 34] this behaviour clearly suggests that contribution of the ionic effects in the charge transport can be considered negligible in HVAE. Moreover, the drastic effects induced by the exposition to soaking[35] water or acidic solutions witness the key role of packing of the aromatic polyindole systems in determining electrical properties of the films.[17, 18, 36]

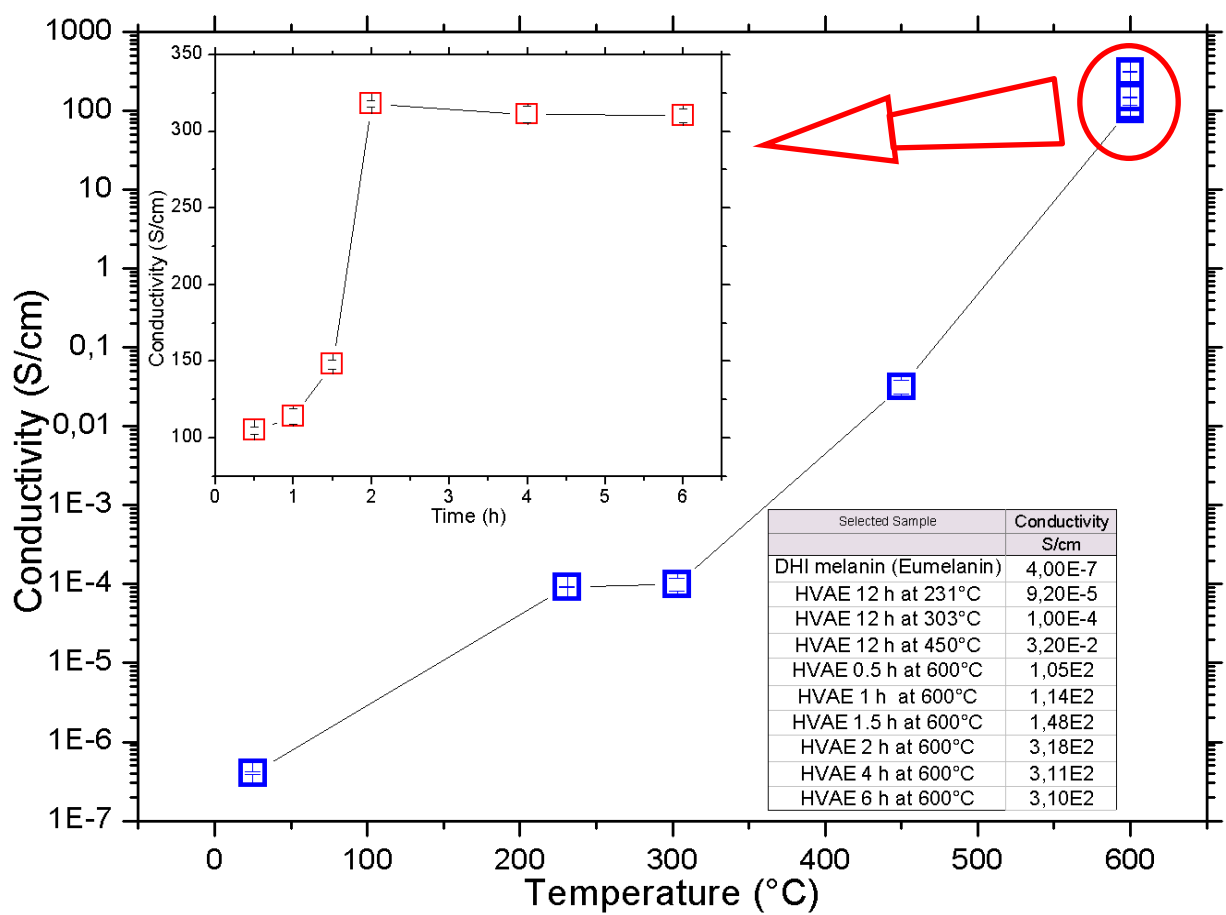

*Figure 4. Conductivity of vacuum annealed eumelanin thin films, vs. the annealing temperature and (inset) vs. the annealing time at 600°C temperature. Data are listed in the table. All the measurements were performed in air at room temperature. Errors of each point are indicated inside the plots symbols.*

The here observed increases in the conductivity cannot be ascribed to the formation of films akin to dense carbon black materials, [37,38] because the processes producing these materials operate at much higher temperature (1000°C or more) when applied to eumelanin-like materials, [11, 12] or anyway using temperatures above 600°C to obtain good conductivity values when applied to polypeptides rich in eumelanin precursors (phenylalanine).[39] Instead, in this study it is observed a conductivity increase from 3 to 5 orders of magnitude even after annealing in the 200°C÷450°C range. This strongly suggests that conductivity rise has not to be ascribed to carbonization processes. Indeed, elemental analysis data (Table S5) do confirm the material does not present C/X ratios expectable for carbon black materials.[37] Measurements of electrical resistance vs. temperature were also performed (Figure S13), using two terminals devices. The observed values of $R$ and the trend of $R$ vs. $T$ reveal that not simple mechanisms are operating for the conductivity of the material: the small values of $R$ indicate that it is a good conductor, while its trend in this range of temperatures cannot discriminate between a nature of semiconductor (decreasing $R$ vs. $T$) or of conductor (increasing $R$ vs. $T$). Nonetheless, at fixed temperature the conductivity appears pretty stable with time (Figure S14), allowing the material to sustain a constant current with a very low increase in the applied voltage along the time, as it can be expected for electronic conductive organics.[40]

## Conclusions

Results here reported indicate a radical modification of the actual picture of the eumelanin charge transport properties, reversing the paradigm according to which eumelanin conductivity increases with the water content of the material. Indeed, if the eumelanin films are rearranged into conductive layers, thanks to a simple thermal annealing in vacuum which succeeds in inducing a structural reorganization of their molecular constituents, the contribution of the electronic current is here demonstrated to be largely preeminent with respect to the ionic one, allowing to obtain unprecedented high conductivity values, up to 318 S/cm in this work, and the mammalian pigment can be considered as an actual conductor.
The conductivity values achieved and their fine tuning allowed by the control of the process conditions, open to possible tailoring of ad-hoc eumelanin-based active layers for a wide range of applications in organic electronics and bioelectronics, deserving further extensive investigations to get a conclusive picture about the conductor vs. semiconductor behaviour of the eumelanin and insights about the mobility of charge carriers.

## Acknowledgements

Authors wish to acknowledge the projects that financed this work: Italian Project RELIGHT (PON02_00556_3306937); the Italian Ministry of Economic Development PROG. No. E10/000798/02/E 17; the European Commission FP7-PEOPLE-2013-IRSES. Project Reference: 612538; KIC EIT RawMaterials Network of Infrastructure OPTNEWOPT (PA15065). Authors thank also Dr Anna De Girolamo Del Mauro for the SEM images, and Dr Carmela Tania Prontera for some of the UV-Vis acquisitions.

**Conflicts of interest**

There are no conflicts to declare.